\documentclass[conference]{IEEEtran}
\IEEEoverridecommandlockouts

\usepackage{cite}
\usepackage{amsmath,amssymb,amsfonts}
\usepackage{algorithmic}
\usepackage{graphicx}
\usepackage{textcomp}
\usepackage{xcolor}
\usepackage[hidelinks,bookmarks=true]{hyperref}
\usepackage[numbered]{bookmark}
\usepackage{soul}
\usepackage{booktabs}
\usepackage{multirow}
\usepackage{float}
\usepackage{url}
\usepackage[small,it]{caption}
\usepackage{tcolorbox}
\usepackage{tabularx} 
\usepackage[switch]{lineno}
\usepackage{titlesec}
\usepackage{pgf-pie}
\usepackage{enumitem}
\usepackage{seqsplit}

\setlist[itemize,enumerate]{noitemsep, topsep=0pt, leftmargin=1.0em}
\hypersetup{
	pdfborder          = {0 0 0},
	breaklinks         = true,
	bookmarksnumbered  = true,
	pdfsubject         = {Tool Demo},
	pdfkeywords        = {program comprehension} {identifier names} {linguistic anti-patterns} {open-source tool},
	pdftitle           = {IDEAL: An Open-Source Identifier Name Appraisal Tool},
	pdfauthor          = {Anthony Peruma, Venera Arnaoudova, Christian D. Newman}
}

\usepackage{array}
\newcolumntype{L}[1]{>{\raggedright\let\newline\\\arraybackslash\hspace{0pt}}m{#1}}
\newcolumntype{C}[1]{>{\centering\let\newline\\\arraybackslash\hspace{0pt}}m{#1}}
\newcolumntype{R}[1]{>{\raggedleft\let\newline\\\arraybackslash\hspace{0pt}}m{#1}}

\begin{document}


\title{\huge IDEAL: An Open-Source Identifier Name Appraisal Tool}

\author{
    \IEEEauthorblockN{Anthony Peruma\IEEEauthorrefmark{1}, Venera Arnaoudova\IEEEauthorrefmark{2}, Christian D. Newman\IEEEauthorrefmark{1}}
    \IEEEauthorblockA{\IEEEauthorrefmark{1}Rochester Institute of Technology, Rochester, NY, USA}
    \IEEEauthorblockA{\IEEEauthorrefmark{2}Washington State University, Pullman, WA, USA}
    axp6201@rit.edu, venera.arnaoudova@wsu.edu, cnewman@se.rit.edu
}

\maketitle

\begin{abstract}
Developers must comprehend the code they will maintain, meaning that the code must be legible and reasonably self-descriptive. Unfortunately, there is still a lack of research and tooling that supports developers in understanding their naming practices; whether the names they choose make sense, whether they are consistent, and whether they convey the information required of them. In this paper, we present IDEAL, a tool that will provide feedback to developers about their identifier naming practices. Among its planned features, it will support linguistic anti-pattern detection, which is what will be discussed in this paper. IDEAL is designed to, and will, be extended to cover further anti-patterns, naming structures, and practices in the near future. IDEAL is open-source and publicly available, with a demo video available at: \url{https://youtu.be/fVoOYGe50zg}
\end{abstract}


\section{Introduction}
Program comprehension is a precursor to all software maintenance task \cite{Rajlich2002WPC}; it is essential that a developer understands the code they will be modifying. Therefore, maintaining the internal quality of the code over its lifetime is of paramount importance. As fundamental elements in the source code, identifier names account, on average, for almost 70\% of the characters in a software system’s codebase \cite{Deissenboeck2006SQJ} and play a significant part in code comprehension \cite{Corbi1989, Martin:2008}. Low quality identifiers can hinder developers' ability to understand the code \cite{Schankin2018ICPC,Lawrie2006ICPC}; well-constructed names can improve comprehension activities by an estimated 19\% \cite{Hofmeister2017SANER}. 

However, there is still very little support for developers in terms of helping them craft high-quality identifier names. Research has examined the terms or structure of names \cite{Hofmeister2017SANER, Deissenboeck2006SQJ, Bonita2010ICPC,Peruma2018IWOR, Abebe2009WCRE} and produced readability metrics and models \cite{Buse2010TSE,Scalabrino2018JSEP,Fakhoury2019ICPC} to try and address this problem. However, they still fall short of providing tangible advice for improving naming practices in developers' day-to-day activities. The work we present in this paper is designed to operate within an IDE, or a CLI, setting and provide real-time advice to developers about their naming practices.

\subsection{Goal}
Our work aims to provide the research and developer community with an open-source tool, IDEAL, that detects and reports violations in identifier names for multiple programming languages using static analysis techniques. In addition to identifying the identifier(s) exhibiting naming issues in the source code, IDEAL also provides necessary information for each reported violation so that appropriate action(s) can be taken to correct the issue. We envision IDEAL utilized by developers in crafting and maintaining high-quality identifier names in their projects and also by the research community to study the distribution and effect that various poor naming practices have in the field.

\subsection{Contribution}
IDEAL is a multi-language platform for identifier name analysis. It is context-aware; treating test and production names differently since they have different characteristics \cite{Newman2020JSS,2021-ICPC-Methods}. It allows for project-specific configurations and is based on srcML \cite{Collard2013ICSM}, allowing it to support multiple programming languages (specifically, Java and C\#). IDEAL is publicly available \cite{ProjectWebsite} as an open-source tool to facilitate extension and use within the researcher and developer communities.


\section{Linguistic Anti-Patterns} 
While the idea behind IDEAL is to support a broad range of identifier naming best practices based on research, we needed a strong place to start fleshing the tool out. We chose to implement the linguistic anti-patterns, which were first conceptualized by Arnaoudova et al. \cite{Arnaoudova2013CSMR}. The primary reasons for this are that the anti-patterns are well-researched and they represent real, tangible identifier naming problems. Further, modern IDEs currently do not support the semantics-aware naming problem detection embodied by the Linguistic Anti-patterns and the current implementations of anti-patterns are: Limited to singular languages, not open source, limited to a single IDE environment, and/or do not provide enough information to the developer to help ameliorate naming issues. Thus, they are a good place for IDEAL to begin providing a direct, positive influence.

Linguistic anti-patterns represent deviations from well-established lexical naming practices in source code and act as indicators of poor naming quality. This degradation in quality  results in inconsistencies in the source code, leading to misinterpretations causing an increase in developer cognitive load \cite{Fakhoury2018ICPC}.  Detecting such naming violations in the source code is typically a tedious and error-prone task for developers that requires an understanding of the system and a manual analysis of the complete source code. Thus, tool support is warranted.


To this extent, studies in linguistic anti-pattern detection investigate the use of static analysis and artificial intelligence (AI) as detection mechanisms. Two variants of LAPD (Linguistic Anti-Pattern Detector) by Arnaoudova et al. \cite{Arnaoudova2013CSMR, Arnaoudova2016EMSE} utilize static analysis to detect these anti-patterns in C++ and Java source code. The C++ version of the tool is available as a standalone command-line executable (but is not open source and not extendable), while the Java version is available as an Eclipse Checkstyle plugin. The authors report an average precision of 72\% for the C++ variant of their tool. Fakhoury et al. \cite{Fakhoury2018SANER} construct and compare AI-based linguistic anti-pattern detection models for Java source code. The authors report F1-Scores of 88.77\% for traditional machine learning models and 74.53\% for deep neural network models. However, these models only report on the presence or absence of a linguistic anti-pattern; details around the type of anti-pattern present are not provided. In contrast, since IDEAL is built on srcML, it supports multiple programming languages. IDEAL also provides finer-grain feedback on the types of anti-patterns present and how to fix them; making it easy to use for developers and researchers. It is also made to be extended with further anti-patterns not supported by prior tools, that have been found through prior research \cite{2021-ICPC-Methods, Newman2020JSS}.

Table \ref{Table:AntiPatterns} summarizes the linguistic anti-patterns currently detected by IDEAL. Anti-Patterns A.* to F.* are the set of original anti-patterns defined by Arnaoudova et al. \cite{Arnaoudova2013CSMR}, while the anti-patterns G.* are anti-patterns unique to IDEAL. Our project website \cite{ProjectWebsite} provides code snippets from real-world open-source systems that highlight examples of these anti-patterns. We should also note that as an open-source tool IDEAL provides the necessary infrastructure for the inclusion of additional anti-patterns.

\begin{table*}
\centering
\caption{Summary of the linguistic anti-pattern detection rules IDEAL utilizes.}
\label{Table:AntiPatterns}
\begin{tabular}{@{}|l|L{3.0cm}|L{14.0cm}|@{}}
\toprule
\multicolumn{1}{|c|}{\textbf{Id}} & \multicolumn{1}{c|}{\textbf{Pattern}}         & \multicolumn{1}{c|}{\textbf{Detection Strategy}}                                                                                                                                                                                                                                                                                  \\ \midrule
A.1                              & ``Get'' more than accessor                   & \begin{tabular}[c]{@{}L{14.0cm}@{}}Impacted Identifiers: Method   Names (excludes test methods)\\      The name starts with `get', the access specifier is public/protected, the   name contains the name of an attribute, the return type is the same as the   attribute type, and the body contains conditional statements\end{tabular} \\ \midrule
A.2                              & ``Is'' returns more than a Boolean           & \begin{tabular}[c]{@{}L{14.0cm}@{}}Impacted Identifiers: Method   Names (excludes test methods)\\      The name starts with a predicate/affirmation related term and the return   type is not boolean\end{tabular}                                                                                                                        \\ \midrule
A.3                              & ``Set'' method returns                       & \begin{tabular}[c]{@{}L{14.0cm}@{}}Impacted Identifiers: Method   Names\\      The name starts with `set' and the return type is not void\end{tabular}                                                                                                                                                                                    \\ \midrule
A.4                              & Expecting but not getting single instance    & \begin{tabular}[c]{@{}L{14.0cm}@{}}Impacted Identifiers: Method   Names (excludes test methods)\\      The last term in the name is singular and the name does not contain terms   that are a collection type and the return type is a collection\end{tabular}                                                                            \\ \midrule
B.1                              & Not implemented condition                    & \begin{tabular}[c]{@{}L{14.0cm}@{}}Impacted Identifiers: Method   Names\\      The name contains conditional related terms in the name or comment and body   does not conditional statements\end{tabular}                                                                                                                                 \\ \midrule
B.2                              & Validation method does not confirm           & \begin{tabular}[c]{@{}L{14.0cm}@{}}Impacted Identifiers: Method   Names (excludes test methods)\\      The name starts with a validation-related term, does not have a return type   and does not throw an exception\end{tabular}                                                                                                         \\ \midrule
B.3                              & ``Get'' method does not return               & \begin{tabular}[c]{@{}L{14.0cm}@{}}Impacted Identifiers: Method   Names (excludes test methods)\\      The name starts with a `get' related term and the return type is void\end{tabular}                                                                                                                                                 \\ \midrule
B.4                              & Not answered question                        & \begin{tabular}[c]{@{}L{14.0cm}@{}}Impacted Identifiers: Method   Names (excludes test methods)\\      The name starts with a    predicate/affirmation related term and the return type is void\end{tabular}                                                                                                                              \\ \midrule
B.5                              & Transform method does not return             & \begin{tabular}[c]{@{}L{14.0cm}@{}}Impacted Identifiers: Method   Names (excludes test methods)\\      The name starts with or an inner term constains a transformation term and   the return type is void\end{tabular}                                                                                                                   \\ \midrule
B.6                              & Expecting but not getting a collection       & \begin{tabular}[c]{@{}L{14.0cm}@{}}Impacted Identifiers: Method   Names (excludes test methods)\\      The name starts with a `get' related term, the name contains a term that is   either plural or a collection type and the return type is not a   collection-based type\end{tabular}                                                 \\ \midrule
C.1                              & Method name and return type are opposite     & \begin{tabular}[c]{@{}L{14.0cm}@{}}Impacted Identifiers: Method   Names (excludes test methods)\\      An antonym relationship exists between terms in an identifiers name and   data type\end{tabular}                                                                                                                                   \\ \midrule
C.2                              & Method signature and comment are opposite    & \begin{tabular}[c]{@{}L{14.0cm}@{}}Impacted Identifiers: Method   Names (excludes test methods)\\      An antonym relationship exists between either terms in an identifiers name   or data type and comments\end{tabular}                                                                                                                \\ \midrule
D.1                              & Says one but contains many                   & \begin{tabular}[c]{@{}L{14.0cm}@{}}Impacted Identifiers:   Attributes, Method Variables and Parameters\\      The last term in the name is singular and the data type is a collection\end{tabular}                                                                                                                                        \\ \midrule
D.2                              & Name suggests Boolean but type does not      & \begin{tabular}[c]{@{}L{14.0cm}@{}}Impacted Identifiers:   Attributes, Method Variables and Parameters\\      The starting term should be predicate/affirmation related and the data type   is not boolean\end{tabular}                                                                                                                   \\ \midrule
E.1                              & Says many but contains one                   & \begin{tabular}[c]{@{}L{14.0cm}@{}}Impacted Identifiers:   Attributes, Method Variables and Parameters\\      The last term in the name is plural and the data type is not a collection\end{tabular}                                                                                                                                      \\ \midrule
F.1                              & Attribute name and type are opposite         & \begin{tabular}[c]{@{}L{14.0cm}@{}}Impacted Identifiers:   Attributes, Method Variables and Parameters\\      An antonym relationship exists between terms in an identifiers name and   data type\end{tabular}                                                                                                                            \\ \midrule
F.2                              & Attribute signature and comment are opposite & \begin{tabular}[c]{@{}L{14.0cm}@{}}Impacted Identifiers:   Attributes, Method Variables and Parameters\\      An antonym relationship exists between either terms in an identifiers name   or data type and comments\end{tabular}                                                                                                         \\ \midrule
G.1                              & Name contains only special characters        & \begin{tabular}[c]{@{}L{14.0cm}@{}}Impacted Identifiers:   Attributes, Method, Method Variables and Parameters\\      The name of the identifier is composed of only non-alphanumeric characters\end{tabular}                                                                                                                             \\ \midrule
G.2                              & Redundant use of ``test'' in method name     & \begin{tabular}[c]{@{}L{14.0cm}@{}}Impacted Identifiers: Methods   (excludes non-test methods)\\      The name starts with the term `test'\end{tabular}                                                                                                                                                                                   \\ \bottomrule
\end{tabular}
\vspace{-4mm}
\end{table*}

\section{IDEAL Architecture}
\begin{figure}[!ht]
 	\centering
 	\includegraphics[scale=0.51]{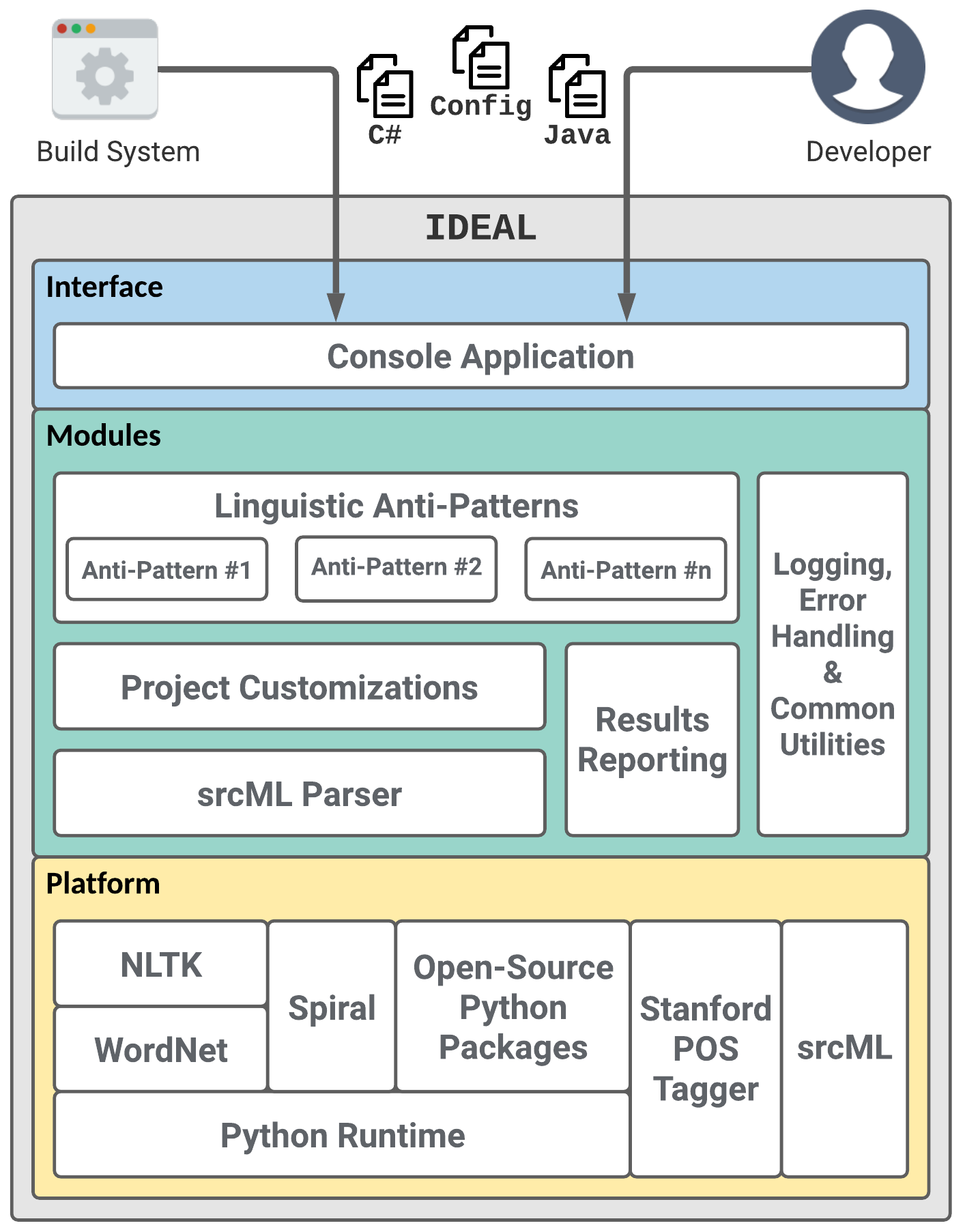}%
  	\vspace{-0.10cm}
 	\caption{Conceptual architectural view of IDEAL.}
 	\label{Figure:architecture}
\vspace{-4mm}
\end{figure}

Implemented as a command-line/console-based tool in Python, IDEAL integrates with some well-known open-source libraries and tools in analyzing source code to detect identifier name violations. Depicted in Figure \ref{Figure:architecture} is a view of the conceptual architecture of IDEAL. Broadly, IDEAL is composed of three layers-- Platform, Modules, and Interface. It utilizes well-known tools and libraries used for natural language and static analysis, including Spiral \cite{Hucka2018JOSS}, NLTK \cite{Bird2009Natural}, Wordnet \cite{Miller1995ACM}, Stanford POS tagging \cite{Toutanova2003HLT}, and srcML \cite{Collard2013ICSM}.

\section{Applicability}
\noindent\textbf{Practitioners.} By integrating IDEAL into their development toolset and workflow, developers are better equipped to maintain identifiers in their source code. As a command-line/console application, the current version of IDEAL supports integration with a build system. Hence, project teams can analyze their entire project codebase, or just what was changed, during their nightly build process and evaluate the report to determine violations that need to be addressed. 

\noindent\textbf{Researchers.} We envision the research community utilizing IDEAL in studies around program comprehension. With the capability of batch-based analysis, IDEAL supports researchers in conducting large-scale empirical studies. Furthermore, by supporting Java and C\#, IDEAL provides researchers to expand their research to multiple programming languages and perform comparatison-based studies. Finally, as an open-source tool, researchers are provided with the opportunity to extend IDEAL by improving existing violation detection strategies and introducing new anti-patterns.

\noindent\textbf{Educators.} IDEAL can be used in a classroom setting to teach students the importance of constructing high-quality identifier names and their impact on software maintenance and evolution. Through this, students will be better prepared to write high-quality code when moving into the industry.


\section{Evaluation}
To understand the effectiveness of IDEAL in correctly detecting identifier naming violations, we subjected IDEAL to two types of evaluation activities. First, we analyzed four popular open-source systems using IDEAL and manually validated the detection results of a statistically significant sample. Our next evaluation strategy involved assessing IDEAL on the sample dataset utilized to evaluate LAPD by comparing the detection results. In the following subsections, we provide details on these two evaluation activities, including numbers around the correctness of IDEAL and qualitative findings based on our manual analysis of source code. 

\subsection{Evaluation on open-source systems}
IDEAL can analyze systems implemented in any language supported by srcML. However, currently, it has only been evaluated using Java and C\#. Thus, we selected two popular open-source systems for each of these programming languages. To this extent, Retrofit \cite{system_retrofit} and Jenkins \cite{system_jenkins} were the two Java systems, while Shadowsocks \cite{system_shadowsocks} and PowerShell \cite{system_powershell} were the C\# systems; Table \ref{Table:Systems} summarizes the release of each system that was part of our evaluation analysis. A breakdown of our validation results is available at \cite{ProjectWebsite}.

For each of the four systems, we manually analyzed a random stratified statistically significant (i.e., confidence level of 95\% and confidence interval of 10\%) set of detected violations for each category. In total, we manually verified 2,019 instances of naming violations spread across the four systems. Table \ref{Table:Validation} provides a breakdown of the number of violation instances for each category. As part of the manual analysis process and to mitigate bias, the authors discussed specific violation instances that were subjective and, at times, referenced literature (grey and reviewed) to aid in the decision-making process. IDEAL reports an average precision of 75.27\%, with 14 out of 19 violation types reporting a precision of over 50\%. Though LAPD reports an average precision of 72\%, we manually validate 1,267 more instances than LAPD. Furthermore, even though IDEAL supports customization per project (e.g., specifying custom collection data types and terms), our evaluation strategy did not utilize this feature in order to maintain consistency in violation detection across the four systems.
\begin{table}
\centering
\caption{Summary of the systems in our evaluation process.}
\label{Table:Systems}
\scriptsize
\begin{tabular}{@{}|llllrr|@{}}
\toprule
\multicolumn{1}{|c}{\textbf{System}} & \multicolumn{1}{c}{\textbf{Language}} & \multicolumn{1}{c}{\textbf{Version}} & \multicolumn{1}{c}{\textbf{\begin{tabular}[c]{@{}c@{}}Release\\Date\end{tabular}}} & \multicolumn{1}{c}{\textbf{\begin{tabular}[c]{@{}c@{}}Files\\Analyzed\end{tabular}}} & \multicolumn{1}{c|}{\textbf{\begin{tabular}[c]{@{}c@{}}Issues\\Detected\end{tabular}}} \\ \midrule
Retrofit & Java & 2.9.0 & May-2020 & 282 & 192 \\ \midrule
Jenkins & Java & 2.293 & May-2021 & 1,688 & 4,818 \\ \midrule
Shadowsocks & C\# & 4.4.0.0 & Dec-2020 & 88 & 275 \\ \midrule
PowerShell & C\# & 7.1.3 & Mar-2021 & 1,290 & 8,455 \\ \bottomrule
\end{tabular}%
\vspace{-5mm}
\end{table}
From Table \ref{Table:Validation}, we observe that while IDEAL performs notably well in detecting all A.*, D.*, and E.* violations (precision score of over 80\%). These are anti-patterns where the identifier either does or contains more than what is required. In most instances, IDEAL can accurately process the return/data type of the identifier to determine violations. However, there are also violations that are challenging for IDEAL to analyze and hence result in a large volume of false positives (e.g., C.2). Our manual analysis of these false-positive instances shows patterns that, in most cases, are causing IDEAL to report them as issues. First, since developers utilize custom data/return types for identifiers in their code, IDEAL fails in identifying their intended purpose. For instance, `EnvVars' is a custom type created by a developer to hold a collection of specific items. The developer returns this type in a method called `getEnvironmentVariables2'. Since IDEAL is unaware that `EnvVars' is a collection-based type, it flags this as a violation since the method is supposed to return a collection (i.e., this get method name contains a plural term-- `Variables'). We are confident that once developers configure IDEAL to handle custom types, false positives, similar to this, will reduce. Our next observation is on how IDEAL analyzes lexical relationships between words; specifically, concerning antonyms (i.e., C.* and F.*). While IDEAL correctly recognizes antonyms, the context around how these terms are used, either in the identifier's name or comment, is not considered, resulting in false positives. Additionally, we also observe that naming habits/conventions also cause the emergence of antonyms. For instance, consider the method `GetCompletionResult' with a return type called `CompletionResult'. IDEAL determines that `Get' and `Result' are antonyms, which are lexically valid, but a false positive due to naming conventions. Similar to the last challenge, context around the use of transformation terms (i.e., B.5) and conditional terms (i.e., B.1) cause the reporting of a high volume of false positives. While IDEAL correctly detects these terms in the source code, how the developer utilizes the term in a name or comment is currently a challenge.

Finally, our manual review of the source code also allowed us to observe other poor naming/coding practices, which can be future linguistic anti-patterns. For example, the generic terms `data' and `result' are subjective. When used as part of an identifier's name, it is unknown if the identifier handles a single item or collection of items. Likewise, the use of the type `var' (in C\#) and `object' also does not indicate the type of data the identifier handles. Ideally, to convey the purpose/behavior of the identifier correctly, developers need to be specific in naming identifiers and data types when possible. 

\begin{table}
\centering
\caption{Summary of the detection correctness of IDEAL.}
\label{Table:Validation}
\scriptsize
\begin{tabular}{@{}|lrrrrr|@{}}
\toprule
\multicolumn{1}{|c}{\textbf{Id.}} & \multicolumn{1}{c}{\textbf{\begin{tabular}[c]{@{}c@{}}Detected\\ Instances\end{tabular}}} & \multicolumn{1}{c}{\textbf{\begin{tabular}[c]{@{}c@{}}Validated\\ Samples\end{tabular}}} & \multicolumn{1}{c}{\textbf{\begin{tabular}[c]{@{}c@{}}True\\ Positives\end{tabular}}} & \multicolumn{1}{c}{\textbf{\begin{tabular}[c]{@{}c@{}}False\\ Positives\end{tabular}}} & \multicolumn{1}{c|}{\textbf{Precision}} \\ \midrule
A.1                               & 53                                                                                        & 34                                                                                       & 34                                                                                    & 0                                                                                      & 100.00\%                                \\ \midrule
A.2                               & 45                                                                                        & 37                                                                                       & 37                                                                                    & 0                                                                                      & 100.00\%                                \\ \midrule
A.3                               & 129                                                                                       & 64                                                                                       & 63                                                                                    & 1                                                                                      & 98.44\%                                 \\ \midrule
A.4                               & 341                                                                                       & 127                                                                                      & 102                                                                                   & 25                                                                                     & 80.31\%                                 \\ \midrule
B.1                               & 912                                                                                       & 171                                                                                      & 73                                                                                    & 98                                                                                     & 42.69\%                                 \\ \midrule
B.2                               & 446                                                                                       & 166                                                                                      & 165                                                                                   & 1                                                                                      & 99.40\%                                 \\ \midrule
B.3                               & 260                                                                                       & 101                                                                                      & 101                                                                                   & 0                                                                                      & 100.00\%                                \\ \midrule
B.4                               & 18                                                                                        & 16                                                                                       & 5                                                                                     & 11                                                                                     & 31.25\%                                 \\ \midrule
B.5                               & 271                                                                                       & 107                                                                                      & 46                                                                                    & 61                                                                                     & 42.99\%                                 \\ \midrule
B.6                               & 827                                                                                       & 159                                                                                      & 128                                                                                   & 31                                                                                     & 80.50\%                                 \\ \midrule
C.1                               & 139                                                                                       & 74                                                                                       & 54                                                                                    & 20                                                                                     & 72.97\%                                 \\ \midrule
C.2                               & 294                                                                                       & 112                                                                                      & 13                                                                                    & 99                                                                                     & 11.61\%                                 \\ \midrule
D.1                               & 3,359                                                                                      & 262                                                                                      & 261                                                                                   & 1                                                                                      & 99.62\%                                 \\ \midrule
D.2                               & 83                                                                                        & 53                                                                                       & 53                                                                                    & 0                                                                                      & 100.00\%                                \\ \midrule
E.1                               & 5,506                                                                                      & 268                                                                                      & 253                                                                                   & 15                                                                                     & 94.40\%                                 \\ \midrule
F.1                               & 38                                                                                        & 32                                                                                       & 19                                                                                    & 13                                                                                     & 59.38\%                                 \\ \midrule
F.2                               & 165                                                                                       & 91                                                                                       & 15                                                                                    & 76                                                                                     & 16.48\%                                 \\ \midrule
G.1                               & 1                                                                                         & 1                                                                                        & 1                                                                                     & 0                                                                                      & 100.00\%                                \\ \midrule
G.2                               & 853                                                                                       & 144                                                                                      & 144                                                                                   & 0                                                                                      & 100.00\%                                \\
\midrule\midrule
\textit{Overall}                               & 13,740                                                                                       & 2,019                                                                                      & 1,567                                                                                   & 452                                                                                      & 75.27\%                                \\
\bottomrule
\end{tabular}%
\vspace{-4mm}
\end{table}

\subsection{Comparison with LAPD}
In this part of our evaluation we compare the correctness of IDEAL with LAPD. To this extent, we analyze a sample of the source files that were utilized to evaluate the effectiveness of LAPD and compare the results. Since IDEAL implements the anti-patterns available in LAPD, it is essential to understand the areas where IDEAL under- and overperforms. In total, we analyzed 209 Java files and detected 294 violations. From this, both IDEAL and LAPD matched 199 true positive instances and 19 false positive instances. Furthermore, 47 instances identified as LAPD false positives were not detected by IDEAL, highlighting where IDEAL outperforms LAPD. Most of these instances were associated with C.2, D.1, and E.1. Finally, we also encounter instances where IDEAL does not detect LAPD true positives. While some of these issues are due to custom data types, we also encounter subjective instances, most of which (10 instances) fall under D.2. 



\section{Conclusion and Future Work}
This paper introduced IDEAL, an open-source configurable tool that detects 19 types of identifier naming violations in Java and C\# code. A comprehensive evaluation of IDEAL reports an average precision of 75.27\%. Our future work involves increasing support of additional anti-patterns and naming structures (including naming structures derived in other research \cite{Newman2020JSS, 2021-ICPC-Methods}), utilizing a source code specialized part-of-speech-tagger \cite{NewmanTSETagger}, and IDE integration. A summary of the naming practices IDEAL will support is available in the Identifier Name Structure Catalogue \cite{Name_Catalogue}. 

\section{Acknowledgements}
This material is based upon work supported by the National Science Foundation under Grant No. 1850412.
\bibliographystyle{ieeetr}
\bibliography{main}

\begin{thebibliography}{10}

\bibitem{Rajlich2002WPC}
V.~{Rajlich} and N.~{Wilde}, ``The role of concepts in program comprehension,''
  in {\em Proceedings 10th International Workshop on Program Comprehension},
  pp.~271--278, 2002.

\bibitem{Deissenboeck2006SQJ}
F.~Deissenboeck and M.~Pizka, ``Concise and consistent naming,'' {\em Software
  Quality Journal}, vol.~14, pp.~261--282, Sep 2006.

\bibitem{Corbi1989}
T.~A. {Corbi}, ``Program understanding: Challenge for the 1990s,'' {\em IBM
  Systems Journal}, vol.~28, no.~2, pp.~294--306, 1989.

\bibitem{Martin:2008}
R.~C. Martin, {\em Clean Code: A Handbook of Agile Software Craftsmanship}.
\newblock Upper Saddle River, NJ, USA: Prentice Hall PTR, 1~ed., 2008.

\bibitem{Schankin2018ICPC}
A.~{Schankin}, A.~{Berger}, D.~V. {Holt}, J.~C. {Hofmeister}, T.~{Riedel}, and
  M.~{Beigl}, ``Descriptive compound identifier names improve source code
  comprehension,'' in {\em 2018 IEEE/ACM 26th International Conference on
  Program Comprehension (ICPC)}, pp.~31--3109, 2018.

\bibitem{Lawrie2006ICPC}
D.~{Lawrie}, C.~{Morrell}, H.~{Feild}, and D.~{Binkley}, ``What's in a name? a
  study of identifiers,'' in {\em 14th IEEE International Conference on Program
  Comprehension (ICPC'06)}, pp.~3--12, 2006.

\bibitem{Hofmeister2017SANER}
J.~{Hofmeister}, J.~{Siegmund}, and D.~V. {Holt}, ``Shorter identifier names
  take longer to comprehend,'' in {\em 2017 IEEE 24th International Conference
  on Software Analysis, Evolution and Reengineering (SANER)}, pp.~217--227,
  2017.

\bibitem{Bonita2010ICPC}
B.~Sharif and J.~I. Maletic, ``An eye tracking study on camelcase and
  under\_score identifier styles,'' in {\em 2010 IEEE 18th International
  Conference on Program Comprehension}, pp.~196--205, 2010.

\bibitem{Peruma2018IWOR}
A.~Peruma, M.~W. Mkaouer, M.~J. Decker, and C.~D. Newman, ``An empirical
  investigation of how and why developers rename identifiers,'' in {\em
  Proceedings of the 2nd International Workshop on Refactoring}, IWoR 2018,
  (New York, NY, USA), p.~26–33, Association for Computing Machinery, 2018.

\bibitem{Abebe2009WCRE}
S.~L. Abebe, S.~Haiduc, P.~Tonella, and A.~Marcus, ``Lexicon bad smells in
  software,'' in {\em 2009 16th Working Conference on Reverse Engineering},
  pp.~95--99, 2009.

\bibitem{Buse2010TSE}
R.~P.~L. {Buse} and W.~R. {Weimer}, ``Learning a metric for code readability,''
  {\em IEEE Transactions on Software Engineering}, vol.~36, no.~4,
  pp.~546--558, 2010.

\bibitem{Scalabrino2018JSEP}
S.~Scalabrino, M.~Linares-Vásquez, R.~Oliveto, and D.~Poshyvanyk, ``A
  comprehensive model for code readability,'' {\em Journal of Software:
  Evolution and Process}, vol.~30, no.~6, p.~e1958, 2018.
\newblock e1958 smr.1958.

\bibitem{Fakhoury2019ICPC}
S.~Fakhoury, D.~Roy, A.~Hassan, and V.~Arnaoudova, ``Improving source code
  readability: Theory and practice,'' in {\em 2019 IEEE/ACM 27th International
  Conference on Program Comprehension (ICPC)}, pp.~2--12, 2019.

\bibitem{Newman2020JSS}
C.~D. Newman, R.~S. AlSuhaibani, M.~J. Decker, A.~Peruma, D.~Kaushik, M.~W.
  Mkaouer, and E.~Hill, ``On the generation, structure, and semantics of
  grammar patterns in source code identifiers,'' {\em Journal of Systems and
  Software}, vol.~170, p.~110740, 2020.

\bibitem{2021-ICPC-Methods}
A.~Peruma, E.~Hu, J.~Chen, E.~A. Alomar, M.~W. Mkaouer, and C.~D. Newman,
  ``Using grammar patterns to interpret test method name evolution,'' in {\em
  Proceedings of the 29th International Conference on Program Comprehension},
  ICPC '21, (New York, NY, USA), Association for Computing Machinery, 2021.

\bibitem{Collard2013ICSM}
M.~L. Collard, M.~J. Decker, and J.~I. Maletic, ``Srcml: An infrastructure for
  the exploration, analysis, and manipulation of source code: A tool
  demonstration,'' in {\em Proceedings of the 2013 IEEE International
  Conference on Software Maintenance}, ICSM '13, (USA), p.~516–519, IEEE
  Computer Society, 2013.

\bibitem{ProjectWebsite}
\url{https://www.scanl.org/artifacts/tools/}.

\bibitem{Arnaoudova2013CSMR}
V.~Arnaoudova, M.~Di~Penta, G.~Antoniol, and Y.-G. Guéhéneuc, ``A new family
  of software anti-patterns: Linguistic anti-patterns,'' in {\em 2013 17th
  European Conference on Software Maintenance and Reengineering}, pp.~187--196,
  2013.

\bibitem{Fakhoury2018ICPC}
S.~Fakhoury, Y.~Ma, V.~Arnaoudova, and O.~Adesope, ``The effect of poor source
  code lexicon and readability on developers' cognitive load,'' in {\em 2018
  IEEE/ACM 26th International Conference on Program Comprehension (ICPC)},
  pp.~286--28610, 2018.

\bibitem{Arnaoudova2016EMSE}
V.~Arnaoudova, M.~Di~Penta, and G.~Antoniol, ``Linguistic antipatterns: what
  they are and how developers perceive them,'' {\em Empirical Software
  Engineering}, vol.~21, pp.~104--158, Feb 2016.

\bibitem{Fakhoury2018SANER}
S.~Fakhoury, V.~Arnaoudova, C.~Noiseux, F.~Khomh, and G.~Antoniol, ``Keep it
  simple: Is deep learning good for linguistic smell detection?,'' in {\em 2018
  IEEE 25th International Conference on Software Analysis, Evolution and
  Reengineering (SANER)}, pp.~602--611, 2018.

\bibitem{Hucka2018JOSS}
M.~Hucka, ``Spiral: splitters for identifiers in source code files,'' {\em
  Journal of Open Source Software}, vol.~3, no.~24, p.~653, 2018.

\bibitem{Bird2009Natural}
S.~Bird, E.~Klein, and E.~Loper, {\em Natural Language Processing with Python:
  Analyzing Text with the Natural Language Toolkit}.
\newblock O'Reilly Media, 2009.

\bibitem{Miller1995ACM}
G.~A. Miller, ``Wordnet: A lexical database for english,'' {\em Commun. ACM},
  vol.~38, p.~39–41, Nov. 1995.

\bibitem{Toutanova2003HLT}
K.~Toutanova, D.~Klein, C.~D. Manning, and Y.~Singer, ``Feature-rich
  part-of-speech tagging with a cyclic dependency network,'' in {\em
  Proceedings of the 2003 Human Language Technology Conference of the North
  American Chapter of the Association for Computational Linguistics},
  pp.~252--259, 2003.

\bibitem{system_retrofit}
\url{https://github.com/square/retrofit}.

\bibitem{system_jenkins}
\url{https://github.com/jenkinsci/jenkins}.

\bibitem{system_shadowsocks}
\url{https://github.com/shadowsocks/shadowsocks-windows}.

\bibitem{system_powershell}
\url{https://github.com/PowerShell/PowerShell}.

\bibitem{NewmanTSETagger}
C.~D. Newman, M.~J. Decker, R.~S. Alsuhaibani, A.~Peruma, M.~W. Mkaouer,
  S.~Mohapatra, T.~Vishoi, M.~Zampieri, T.~J. Sheldon, and H.~Emily, ``An
  ensemble approach for annotating source code identifiers with part-of-speech
  tags,'' {\em IEEE Transactions on Software Engineering}, 2021.

\bibitem{Name_Catalogue}
\url{https://github.com/SCANL/identifier_name_structure_catalogue}.

\end{thebibliography}

\end{document}